\documentclass[conf]{new-aiaa}
\usepackage[utf8]{inputenc}
\usepackage{graphicx}
\usepackage{amsmath}
\usepackage[version=4]{mhchem}
\usepackage{siunitx}
\usepackage{longtable,tabularx}
\usepackage{xcolor}

\usepackage{fancyhdr}
\usepackage{blindtext}
\usepackage{eso-pic, rotating, graphicx}

\title{Hypersonic Boundary Layer Transition and Heat Loading}
\author{Ahmad Payvan, Luis Bravo, Anindya Ghoshal, Olaf Marxen, George Karniadakis} 

\affil{Applied Mathematics Department, Brown University, Providence, Rhode Island, USA}
\affil{School of Mechanical Engineering Sciences, University of Surrey, Guildford, UK}
\affil{DEVCOM Army Research Laboratory, Aberdeen Proving Ground, MD, USA}

\begin{document}
\maketitle

\begin{abstract}

Hypersonic boundary layer transition using high-order methods for direct numerical simulations (DNS) is largely unexplored, although a few references exist in the literature. Experimental data in the hypersonic regime are scarce, while almost all existing hypersonic codes have low-order accuracy, which could lead to erroneous results in long-time integration and at high Reynolds numbers. Here, we focus on the transition from laminar to turbulent flow, where the Nusselt number may be five times or higher than the Nusselt number in the turbulent regime. The hypersonic flow regime must be modeled accurately using realistic chemistry to predict heat flux on the surface correctly. In this study, we simulate hypersonic boundary layer transition on a flat plate to compute the thermal and shear stresses on the wall. The domain is initialized with a laminar Blasius solution of compressible boundary layer equations. The transition to turbulence is induced using suction and blowing from a thin strip on the wall. Sponge regions are defined at the inflow and outflow boundaries to eliminate the solution contamination due to the reflections of the boundary conditions. In the spanwise direction, periodic boundary conditions are imposed. The numerical method applied here is a sixth-order compact finite difference in the interior of the domain coupled with a fourth-order Runge-Kutta time-stepping scheme for a structured Cartesian grid using staggered variables. The base finite difference code can perform simulations of non-equilibrium hypersonic flows. The numerical scheme is stabilized using the approximate deconvolution model (ADM) and artificial diffusion coefficients. We investigated the effect of non-equilibrium chemistry on the non-linear instability growth with hypersonic boundary layers. We also quantify thermal loads  during the boundary layer laminar to turbulent transition for Mach number 10. 

\end{abstract}

\section{Introduction}

The transition from laminar to turbulent flow significantly amplifies heat loads and skin friction at hypersonic Mach numbers. Identifying the boundary-layer transition location is crucial for developing effective thermal protection systems. Despite years of research, the transition process in hypersonic boundary layers still needs to be better understood. This lack of comprehension regarding the physics of transition poses a significant challenge to developing reliable turbulence models. Consequently, practical applications such as reusable launch vehicles, high-speed interceptors, and hypersonic cruise vehicles are hindered by this limitation.

The air molecules go through dissociation and recombination chemical reactions at hypersonic speeds. The air composition then changes, and non-equilibrium chemistry effects become significant. Non-equilibrium chemistry could profoundly change the aerodynamic behavior of the air and transition modes. Most direct numerical simulations of hypersonic boundary layer transition have been performed using equilibrium or frozen chemistry assumptions \cite{sivasubramanian2011transition, egorov2006direct}. Linear and non-linear instability growth was investigated in a two-dimensional setup using a second-order finite volume technique for Mach 6 flow over a flat plate \cite{egorov2006direct}. The air was modeled using the ideal gas assumption with constant specific heat. Sivasubramanian and Fasel \cite{sivasubramanian2011transition} employed ideal gas with constant specific heat coefficients to investigate linear and non-linear instability growth during the transition of the boundary layer over a flat plate at Mach 6. They observed that the induced suction/blowing pulse develops into a wave packet that goes through linear and non-linear regimes. Marxen \emph{et al.} \cite{marxen2014direct} employed frozen gas and non-equilibrium chemistry assumptions to investigate weakly non-linear instability in large-amplitude two-dimensional waves. In their study, the hypersonic flow over a flat plate was at Mach number 10, and the Reynolds number was $10^5$. The wall boundary condition in their study was adiabatic, which enforces zero heat flux into the wall. Therefore, they could not investigate the effect of real chemistry on the heat flux. In another study, Rezno and Urzay \cite{di2021direct} conducted DNS of Mach 10 hypersonic boundary layer transition with non-equilibrium chemistry and reported a thorough statistical analysis of flow quantities for the flow over an isothermal cold wall. In the current study, we investigate the effect of transition on heat load over a hot flat plate. 

In this study, we study the effect of employing real chemistry versus frozen chemistry on wave packet development in hypersonic boundary layer transition at Mach 10. This study employs a sixth-order accurate compact finite difference scheme to reduce unnecessary numerical dissipation to capture the transition to turbulence. We have also performed simulations of a three-dimensional hypersonic boundary layer transition to observe non-linear instability growth into turbulent regimes and investigate the effect of transition on the heat load over hypersonic vehicles. The article is organized as follows. First, we present the computational methodology and the governing equations. Next, we discuss the results of the 2D  and 3D boundary layer transitions. At the end, we conclude and present a summary of our findings.


\section{Computational Methodology}

\subsection{Direct Numerical Simulation (DNS)}
Here, we solve the two and three-dimensional reactive and non-reactive compressible Navier-Stokes equations. In this section, we describe the Navier-Stokes equations that are formulated for a gas mixture \cite{marxen2013method} as

\begin{equation}
\frac{\partial \rho}{\partial t}+\frac{\partial}{\partial x_j} \left(\rho u_j\right)=0,
    \label{conv_rho}
\end{equation}
\begin{equation}
\frac{\partial \rho^k}{\partial t}+\frac{\partial}{\partial x_j} \left(\rho^k u_j+ J_j^k\right)=\dot{w}^k, \quad k=1,\cdots,N_s,
    \label{conv_rhos}
\end{equation}
\begin{equation}
\frac{\partial (\rho u_i)}{\partial t}+\frac{\partial}{\partial x_j} \left(\rho u_i u_j+ p\delta_{ij}\right)=\frac{\partial  \sigma_{ij}}{\partial x_j}, \quad i=1,2,3, 
    \label{momnetum}
\end{equation}
\begin{equation}
\frac{\partial E}{\partial t}+\frac{\partial}{\partial x_j} \left[(E+p) u_j\right]=-\frac{\partial q_j}{\partial x_j}+\frac{\partial ( u_j\sigma_{ij})}{\partial x_j}.
    \label{energy}
\end{equation}
In Eqs.~\eqref{conv_rho}-\eqref{energy}, $u_j$ denotes the velocity in $j$ direction, $\rho$ is density,and  $\rho^k$ is the density of $k^\textrm{th}$ species. We now define the viscous stress $\sigma_{ij}$, the heat flux vector $q_j$, and the total energy per unit volume $E$ as
\begin{equation}
\sigma_{ij}=\frac{\mu}{Re_\infty}\left(\frac{\partial u_i}{\partial x_j}+\frac{\partial u_j}{\partial x_i}-\frac{2}{3}\frac{\partial u_k}{\partial x_k}\delta_{ij}\right),
    \label{viscous}
\end{equation}
\begin{equation}
q_j=-\frac{1}{Re_\infty Pr_\infty Ec_\infty}\kappa\frac{\partial T}{\partial x_j}+\sum_{k=1}^{N_s}h^k J^k_j,
    \label{heat_flux}
\end{equation}
\begin{equation}
E=\frac{1}{Ec_\infty}\rho e+\frac{1}{2}\rho u_i u_i,
    \label{heat_flux}
\end{equation}
where $\mu$ denotes the mixture viscosity, $\kappa$ indicates the mixture thermal conductivity, $h^k$ is the enthalpy for species $k$ and $J_j^k$ is the mass diffusion flux for species $k$. The Reynolds number $Re_\infty$, the Prandtl number $Pr_\infty$, and the Eckert number $Ec_\infty$ are all defined in \cite{marxen2013method}. The non-dimensionalization of the equations is described in \cite{marxen2013method}. We do not consider the species transport equations and mass diffusion fluxes when we employ frozen chemistry. The non-equilibrium chemistry model considers five species including $O_2$, $N_2$, $NO$, $N$, and $O$ in the gas mixture. The transport properties of the gas mixture are computed using the approach of \cite{MAGIN2004424}.

\subsection{Numerical methods}
The numerical scheme used in this study is a compact sixth-order finite difference along with a fourth-order Runge-Kutta time-stepping scheme for temporal discretization. The numerical approach is based on the numerical algorithm described in \cite{NAGARAJAN2003392} to solve non-dimensional equations of motion for a calorically perfect gas. The numerical solver employs a Cartesian staggered grid to discretize the equations of motion. Marxen \emph{et al.} \cite{marxen2013method} extended the method to solve reactive Navier-Stokes equations with finite-rate chemistry.

\subsection{Simulation setup}
A hypersonic flow with Mach number $M=10$ and Reynolds number $Re=10^5$ is considered over a flat plate. The physical domain consists of a cuboid for which $x\in[11.61,154.41]$, $y\in[0,1.575]$, and $z\in[-\lambda/2,\lambda/2]$. The term $\lambda=2\pi/\beta$ denotes the spanwise length of the physical domain and $\beta=0.8$. A constant temperature of $T_{wall}=\alpha T_\infty$ is imposed at the iso-thermal wall. The suction/blowing perturbations are applied by defining a time-dependent oscillatory profile for the normal momentum over a thin strip on the wall as 

\begin{equation}
(\rho v)_{wall}=\sum_{k=0}^6 A_{v,k}\sin(2\pi \xi-\omega t +\theta_k)\cos(k\beta z)e^{-4\xi^2},\quad \xi=\frac{x-x_{s,1}}{L_s}
    \label{sucblow}
\end{equation}
(U) where $A_{v,0}=0.025M$ and $A_{v,k}=0.01M$ for $k=1,\cdots,6$, and $\theta_0=0.0$ and $\theta_k=\pi/4$ for $k=1,\cdots,6$. The non-dimensional frequency $\omega=34$ is selected to trigger nonlinear instabilities leading to a transition to turbulence downstream. For the 2D cases, the span-wise dependent terms will be omitted in Eq.~\eqref{sucblow}. The amplification magnitudes are set at $A_{v,0}=0.006M$ and $A_{v,k}=0.0025M$. 

\begin{figure}[htb]
\begin{center}
\includegraphics[width=0.9\textwidth]{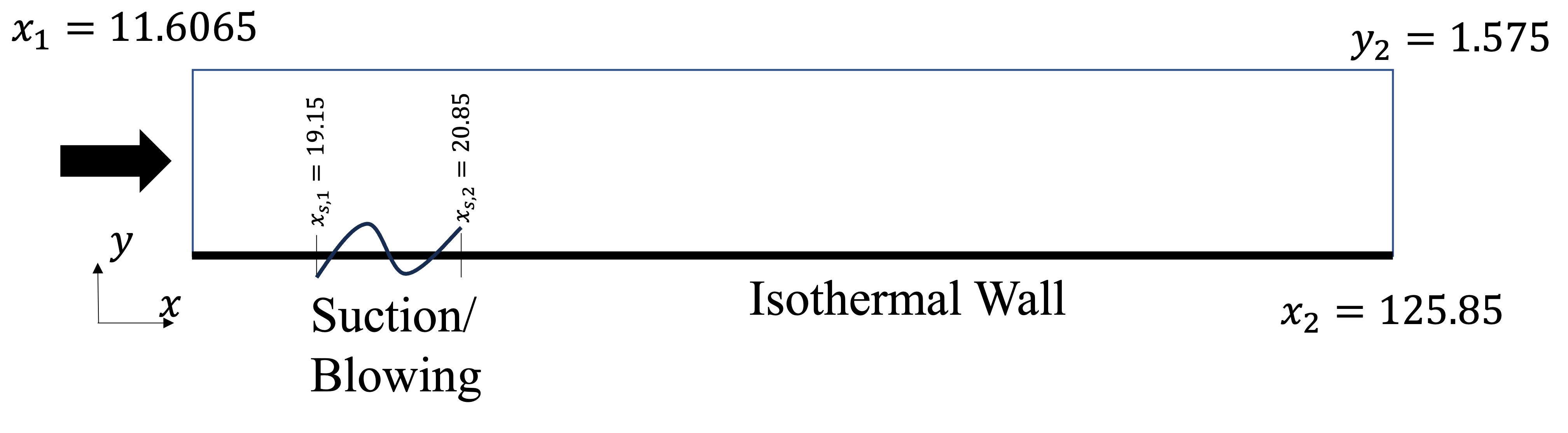}
\end{center}
\caption{(U)  Two-dimensional physical domain schematic. The suction and blowing section is located in a narrow strip on the isothermal wall.
}
\label{fig:2Dschem}
\end{figure}



\section{Results and Discussions}
In this section, we first compare employing non-equilibrium and frozen chemistry kinetics in a 2D configuration boundary layer problem. We then focus on evolution of induced instabilities in a 3D domain where the structure breakdown occurs. 

\subsection{Comparing real versus ideal chemistry kinetics}
We designed a 2D hypersonic boundary layer setup such that a rectangular physical domain with $x\in[11.61,154.41]$ and $y\in[0,1.575]$ was initialized with the solution of compressible laminar boundary layer problem for a flow with $M=10$, $Re_\infty=10^5$, $Pr_\infty=0.69$, $\gamma_\infty=1.397$, $T_\infty=350$K, $\rho_\infty=0.03565 kg/m^3$, and $p_\infty=3596$ Pa. A schematic of the physical domain is shown in Fig.~\ref{fig:2Dschem}. The physical domain is discretized using $N_x=1000$ and $N_y=101$ grid points. In $y-$direction, we have applied grid stretching according to the work of Marxen \emph{et al.} \cite{marxen2013method}.

\begin{figure}[t!]
\begin{center}
\begin{tabular}{cc}
 \includegraphics[width=0.48\textwidth]{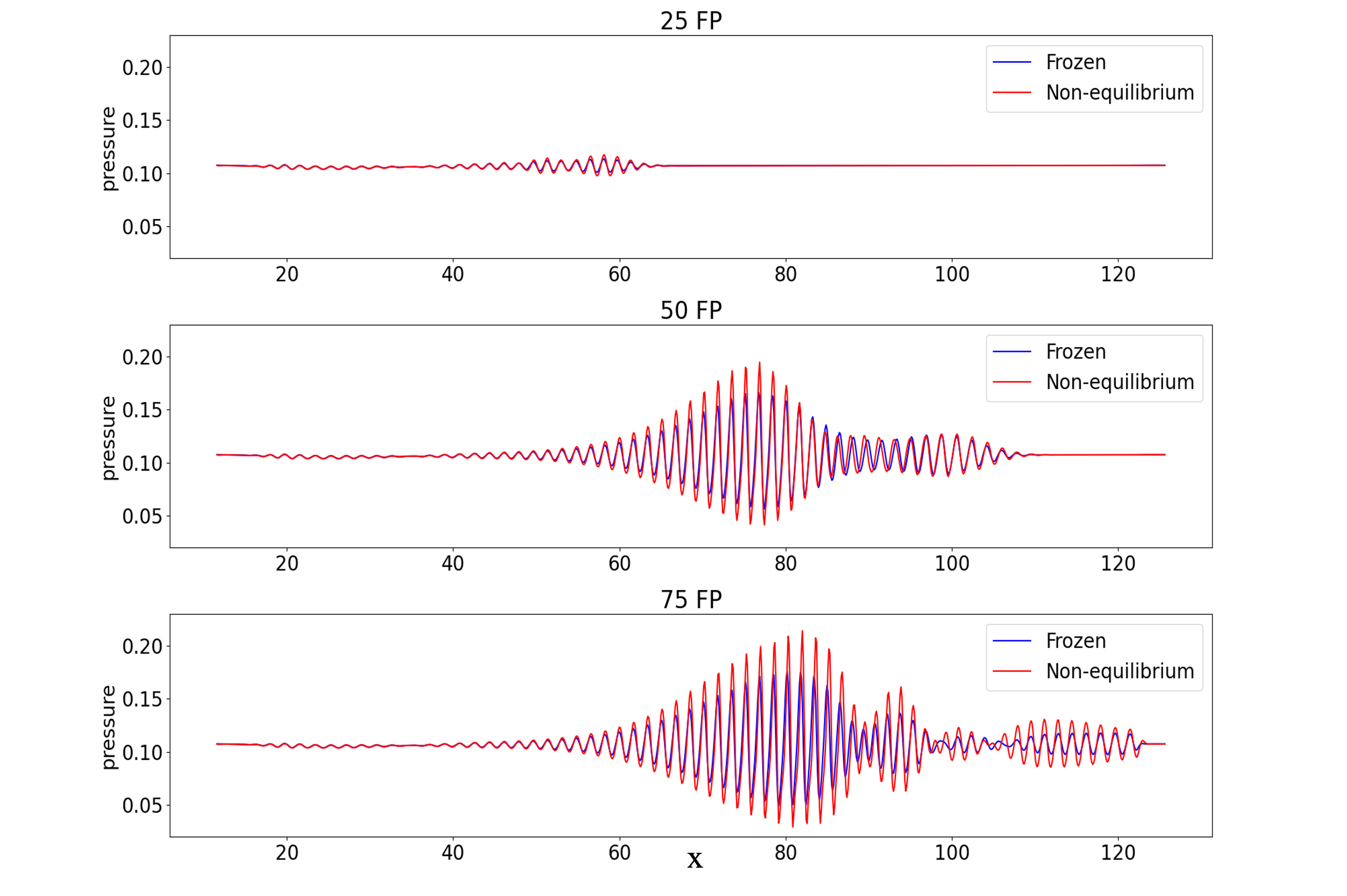}    & \includegraphics[width=0.48\textwidth]{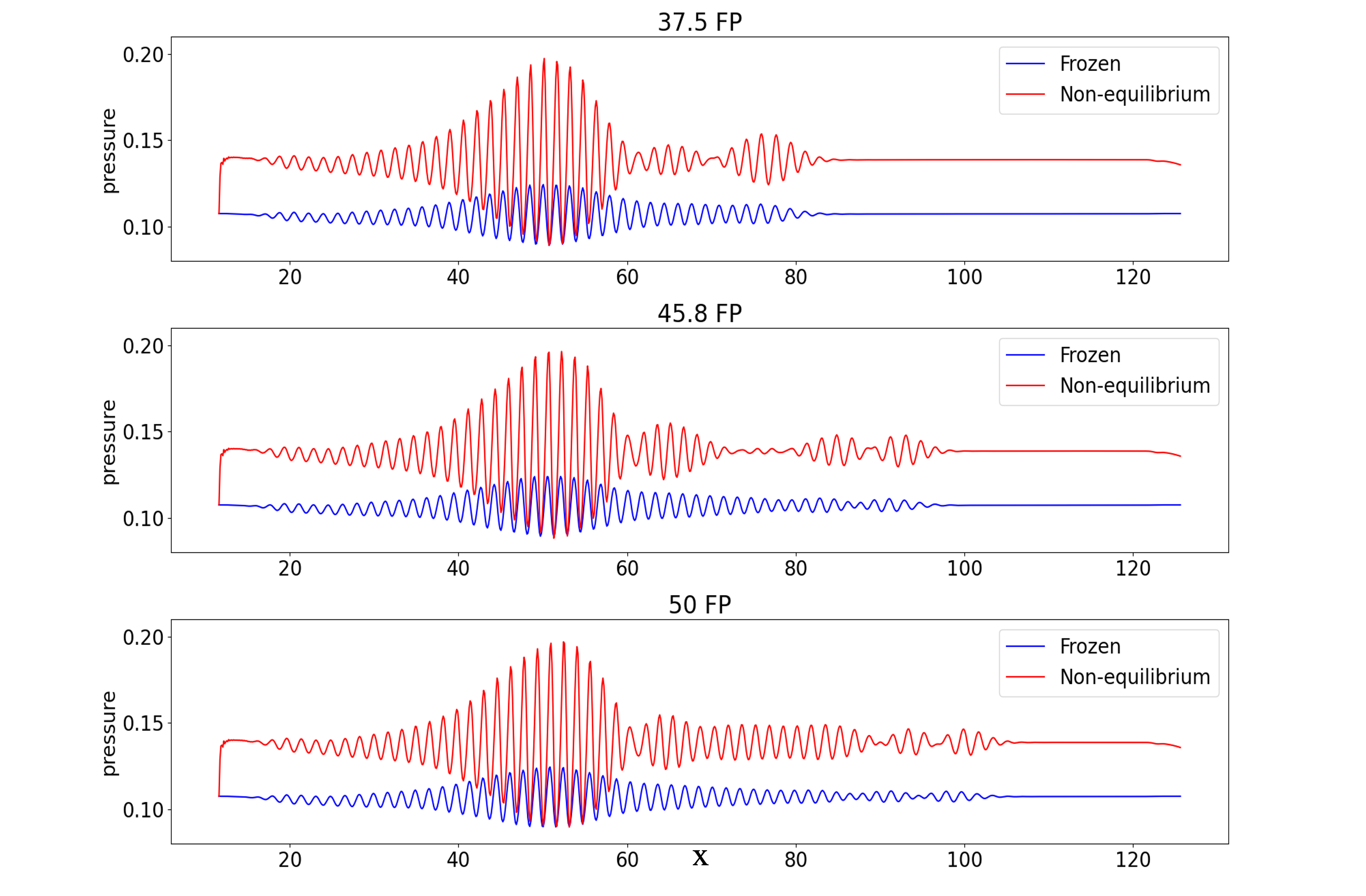} \\ 
 (a) $T_w=4.5T_\infty$ & (b) $T_w=7.0T_\infty$
\end{tabular}
\end{center}
\caption{(U) Pressure distribution along the flow direction: Comparison of  real chemistry versus ideal chemistry for (a) $T_{w}=4.5T_\infty$, the forcing frequency $\omega=34$ and (b) $T_{w}=7.0T_\infty$, the forcing frequency $\omega=36$.
}
\label{fig:realvsideal}
\end{figure}

Figure~\ref{fig:realvsideal}(a) and (b) depict the pressure wave packets for two 2D non-linear instability growth within the hypersonic boundary layer. The pressure profile on the iso-thermal wall is plotted for $T_w=4.5T_\infty$ and $T_w=7.0T_\infty$ cases. For each wall temperature, frozen and non-equilibrium chemistry is considered to reveal the effect of real chemistry on the route to transition. According to Fig.~\ref{fig:realvsideal}(a), the disturbances induced by suction and blowing grow linearly, shown by the $t=25$ forcing period (FP) plot, first, then the growth takes a non-linear form farther downstream. At the stream-wise location $x=80$, the highest amplification occurs, and then wave packets separate from this amplification point and get suppressed towards the end of the domain. For the $ T_w=4.5T_\infty $ case, the pressure waves obtained by the non-equilibrium chemistry obtain the same phase shift as the frozen chemistry case. However, the amplitude of amplifications for the non-equilibrium chemistry case is higher than the frozen one. Considering Fig.~\ref{fig:realvsideal}(b), the effect of non-equilibrium chemistry on the instability growth is profound when $T_w=7.0T_\infty$. Figure~\ref{fig:realvsideal} (b) exhibits pressure profiles on the isothermal wall at three specific time values measured as $t=37.5$, $45.8$, and $50$ FP. We can observe that for the non-equilibrium case, the initial disturbances can induce a wider range of frequencies downstream, meaning a faster transition to turbulence and a higher amplification of the initial disturbances than the frozen chemistry model.
\subsection{Heat load in three dimensional boundary layer transition }
\begin{figure}[t!]
\begin{center}
\includegraphics[width=0.9\textwidth]{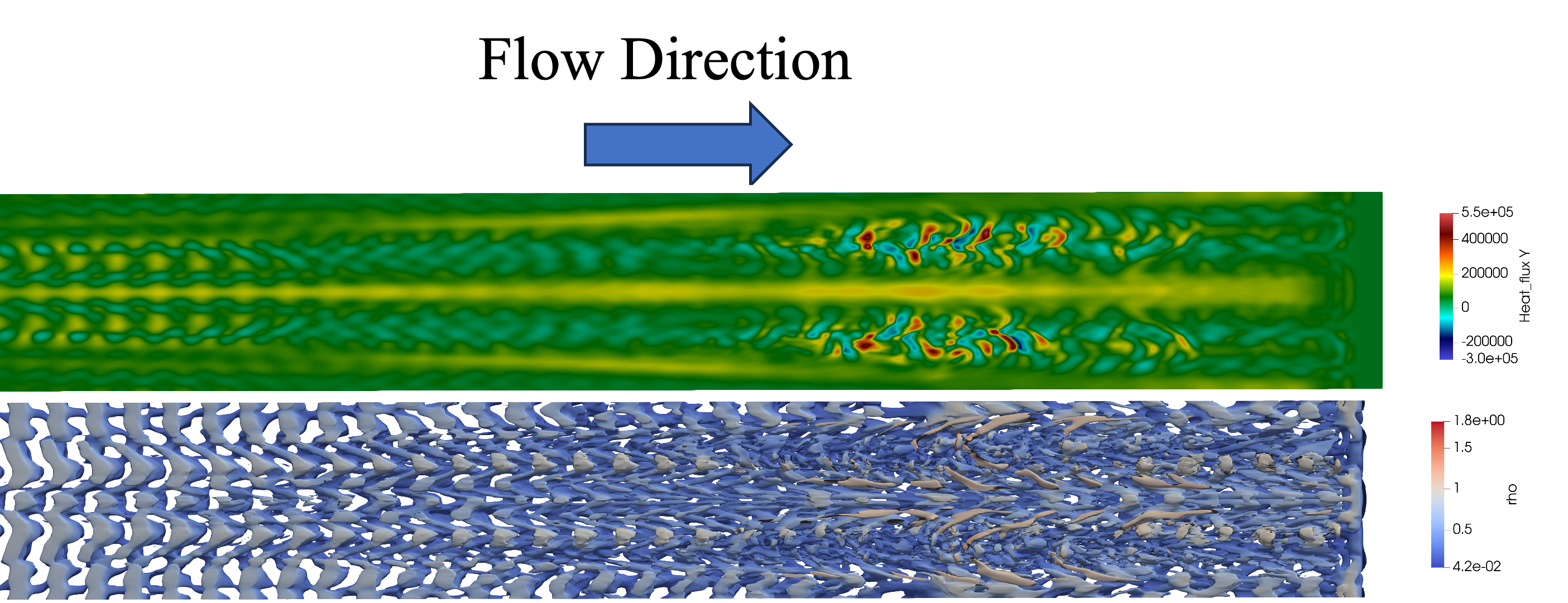}
\end{center}
\caption{Top figure: Contours of heat flux on the wall in $W/m^2$ , Bottom figure: Iso-surfaces of Q-criterion=0.01 colored by density magnitude for the 3D hypersonic boundary layer transition case.
}
\label{fig:3dbl}
\end{figure}

The 3D boundary layer simulation is performed for $T_w=4.5T_\infty$ on a cuboid physical domain which is discretized using $N_x=2000$, $N_y=101$, and $N_z=128$ points in x, y, and z directions, respectively. The numerical simulation is stabilized employing an approximate deconvolution model (ADM) \cite{stolz2001approximate} and artificial diffusion coefficients \cite{kawai2010assessment}. The simulation started using a coarser grid, and then after several FPs, the solution was interpolated up to a higher resolution to save computational time. Figure~\ref{fig:3dbl} shows heat flux contours on the isothermal wall and iso-surfaces of Q-criterion=0.01 colored by the density value. The Q-criterion plot shows that the laminar to turbulent transition occurs as the flow loses symmetry and structure breakdown appears. According to Figure~\ref{fig:3dbl} heat flux map, the transition to turbulence can increase the heat load drastically. The heat load amplification in the transition region shows a higher value than the turbulent region, indicating the significance of modeling transitional regimes for designing hypersonic vehicles.

\section{Summary and Conclusions}

In this research, we utilized a sixth-order compact finite difference method to simulate hypersonic boundary layer transition on a flat plate. Initially, a 2D configuration was employed to compare the impact of real chemistry versus ideal chemistry modeling on hypersonic flows. The simulations were conducted under varied isothermal wall temperatures, revealing that elevated wall temperatures intensify non-equilibrium effects, leading to increased wave growth and fast turbulence transition. Subsequently, 3D simulations of hypersonic boundary layer transition highlighted a significant rise in surface heat load within the transition zone. Consequently, simulating accurate the transition process using a high-order method is crucial in the design phase of hypersonic vehicles.


\section*{Acknowledgments}
This research was supported by DEVCOM Army Research Laboratory award W911NF-12-1-0222 through a Congressional add-on and part of the ARO Modeling of Complex Systems Program. Dr. Olaf Marxen from the University of Surrey is acknowledged for providing the initial computational solver. The authors gratefully acknowledge the High-Performance Computing Modernization Program (HPCMP) resources and support provided by the Department of Defense Supercomputing Resource Center (DSRC) as part of the 2022 Frontier Project, Large-Scale Integrated Simulations of Transient Aerothermodynamics in Gas Turbine Engines. The views and conclusions contained in this document are those of the authors and should not be interpreted as representing the official policies or positions, either expressed or implied, of the DEVCOM Army Research Laboratory or the U.S. Government. The U.S. Government is authorized to reproduce and distribute reprints for Government purposes notwithstanding any copyright notation herein.

\bibliography{refs/sample.bib}

\begin{thebibliography}{9}
\newcommand{\enquote}[1]{``#1''}
\providecommand{\natexlab}[1]{#1}
\providecommand{\url}[1]{\texttt{#1}}
\providecommand{\urlprefix}{URL }
\expandafter\ifx\csname urlstyle\endcsname\relax
  \providecommand{\doi}[1]{\discretionary{}{}{}https://doi.org/#1}\else
  \providecommand{\doi}[1]{\discretionary{}{}{}\urlstyle{rm}\url{https://doi.org/#1}}\fi

\bibitem[{Sivasubramanian and Fasel(2011)}]{sivasubramanian2011transition}
Sivasubramanian, J., and Fasel, H., \enquote{Transition initiated by a localized disturbance in a hypersonic flat-plate boundary layer,} \emph{49th AIAA Aerospace Sciences Meeting including the New Horizons Forum and Aerospace Exposition}, 2011, p. 374.

\bibitem[{Egorov et~al.(2006)Egorov, Fedorov, and Soudakov}]{egorov2006direct}
Egorov, I., Fedorov, A., and Soudakov, V., \enquote{Direct numerical simulation of disturbances generated by periodic suction-blowing in a hypersonic boundary layer,} \emph{Theoretical and Computational Fluid Dynamics}, Vol.~20, 2006, pp. 41--54.

\bibitem[{Marxen et~al.(2014)Marxen, Iaccarino, and Magin}]{marxen2014direct}
Marxen, O., Iaccarino, G., and Magin, T.~E., \enquote{Direct numerical simulations of hypersonic boundary-layer transition with finite-rate chemistry,} \emph{Journal of Fluid Mechanics}, Vol. 755, 2014, pp. 35--49.

\bibitem[{Di~Renzo and Urzay(2021)}]{di2021direct}
Di~Renzo, M., and Urzay, J., \enquote{Direct numerical simulation of a hypersonic transitional boundary layer at suborbital enthalpies,} \emph{Journal of Fluid Mechanics}, Vol. 912, 2021, p. A29.

\bibitem[{Marxen et~al.(2013)Marxen, Magin, Shaqfeh, and Iaccarino}]{marxen2013method}
Marxen, O., Magin, T.~E., Shaqfeh, E.~S., and Iaccarino, G., \enquote{A method for the direct numerical simulation of hypersonic boundary-layer instability with finite-rate chemistry,} \emph{Journal of Computational Physics}, Vol. 255, 2013, pp. 572--589.

\bibitem[{Magin and Degrez(2004)}]{MAGIN2004424}
Magin, T.~E., and Degrez, G., \enquote{Transport algorithms for partially ionized and unmagnetized plasmas,} \emph{Journal of Computational Physics}, Vol. 198, No.~2, 2004, pp. 424--449.

\bibitem[{Nagarajan et~al.(2003)Nagarajan, Lele, and Ferziger}]{NAGARAJAN2003392}
Nagarajan, S., Lele, S.~K., and Ferziger, J.~H., \enquote{A robust high-order compact method for large eddy simulation,} \emph{Journal of Computational Physics}, Vol. 191, No.~2, 2003, pp. 392--419.

\bibitem[{Stolz et~al.(2001)Stolz, Adams, and Kleiser}]{stolz2001approximate}
Stolz, S., Adams, N.~A., and Kleiser, L., \enquote{The approximate deconvolution model for large-eddy simulations of compressible flows and its application to shock-turbulent-boundary-layer interaction,} \emph{Physics of Fluids}, Vol.~13, No.~10, 2001, pp. 2985--3001.

\bibitem[{Kawai et~al.(2010)Kawai, Shankar, and Lele}]{kawai2010assessment}
Kawai, S., Shankar, S.~K., and Lele, S.~K., \enquote{Assessment of localized artificial diffusivity scheme for large-eddy simulation of compressible turbulent flows,} \emph{Journal of computational physics}, Vol. 229, No.~5, 2010, pp. 1739--1762.

\end{thebibliography}

\end{document}